\documentclass[showpacs,amsmath,amssymb,showkeys,pre]{revtex4}
\usepackage{graphicx}% Include figure files
\usepackage{dcolumn}% Align table columns on decimal point
\usepackage{bm}% bold math
\begin{document}
\newcommand{\beq}{\begin{equation}}
\newcommand{\eeq}{\end{equation}}
\newcommand{\beqa}{\begin{eqnarray}}
\newcommand{\eeqa}{\end{eqnarray}}
\newcommand{\non}{\nonumber}
%\preprint{}
\title{The Marginal Stability of the Metastable  TAP States.}
\author{ T.Plefka}
\email[email:]{plefka@web.de}
%\thanks{}
\affiliation{Institute of Condensed Matter Physics, TU Darmstadt, D 64289 Darmstadt, Germany}

\date{\today}
\begin{abstract}
The existing investigations on the complexity are extended. In addition  to the Edward-Anderson Parameter $ q_2$ the fourth moment $ q_4= 1/N \sum_i m_i^4$
of the magnetizations $ m_i $ is included to the set  of constrained variables and the  constrained complexity  $\Sigma(T;q_2,q_4)$ is numerical determined. The maximum of $\Sigma(T;q_2,q_4)$ (representing the total complexity) sticks at the boundary  for temperatures at and below a new critical temperature. This implies  marginal stability for the nearly all metastable states. The temperature dependence of the lowest value of the Gibbs potential consistent with various physical requirements is presented.
\end{abstract}

\pacs{75.10.Nr, 05.50.+q, 87.10.+e}
\keywords{Spin-Glasses, TAP Equations, Complexity}

\maketitle

\section{Introduction}
 The Thouless-Anderson-Palmer (TAP) approach  \cite{tap,tp,tt} to the Sherrington and Kirkpatrick  (SK) model \cite{sk} plays a central role in the attempt to understand the physics of spin glasses and related interdisciplinary problems (neural networks, computer science, theoretical biology, econophysics etc.). The system exhibits meta-stability below the spin glass temperature.
 
  The well established work of Bray and Moore (BM) \cite{bm}  
leads to a finite complexity below the spin glass temperature, 
which implies an exponential increase of the number of metastable TAP states  with increasing system size. It is essential  to count exclusively the physical
states and neglect non-physical ones. However, the BM and subsequent work \cite{CGPM,abm,muller} do not  completely satisfy  this requirement. 

An alternative method to work out the characteristic properties of
the metastable TAP states are  numerical investigations \cite{BM79,mod2,p03,abm,am19} based on iteration techniques or phenomenological dynamics for systems of finite size.  The numerical determined fix-points are interpreted as metastable  TAP states.
Extrapolation to  infinite size systems results in the conclusion
that these states are marginally  stable. Such an extrapolation procedure, however, leads  generally to some uncertainty. Note in this context,  that the numerical investigation are usually performed for systems with just some hundreds spins. Increasing  the system size results in a drastic reduce of the rate  finding  a TAP state. 

In this work the existing approaches on the complexity are extended by the additional inclusion of the forth moment of the magnetizations to the set of constrained variables. This procedure enables a complete 
counting of the physical TAP states. In section II the results of the calculation are presented. The total complexity and various averages are  discussed to some detail in section III. Finally  conclusions are drawn in section IV.

\section{calculation}

More than four decades ago  SK introduced   the spin glass Hamiltonian 
$$ H= -\frac{1}{2} \sum_{i\neq j} J_{i j} S_i S_j
$$
 of $ N $ Ising spins
($S_i=\pm1)$. The bonds $ J_{ij}$ are independent random variables with zero
means and standard deviations $ N^{-1/2}$  (which fixes the spin glass
temperature to $T_{sg}=1$).
 According to the TAP  approach \cite{tap,tp,tt}  the energy
\begin{equation}\label{t3}
U= N (w -\frac{1}{2 T} (1-q_2)^2 )  \quad, \quad w=-\frac{1}{2 N}\sum_{i\neq j}J_{ij}m_i m_j \;,
\end{equation}
 the entropy 
\begin{equation}\label{t4}
 S= \sum_i  s_0 (m_i) -\frac{N}{4 T^2} (1-q_2)^2 \quad,\quad s_0(m)=-\frac{1+m}{2} \ln\frac{1+m}{2}-\frac{1-m}{2} \ln \frac{1-m}{2} 
\end{equation}
and consequently  the Gibbs potential $ G(T,m_i)= U - T S $ 
are given in terms of  the local magnetizations $ m_i$ and the temperature $T$  , where $ q_k$  is defined as
\begin{equation}\label{qq}
 q_k=N^{-1}\sum_i m_i^k \quad (k= 2, 4).
\end{equation}
In general the local magnetic fields $ h_i $ are determined by the  TAP equations  $ h_i= \partial G /\partial m_i $.  This work is exclusively restricted to the case $ h_i=0$  and the TAP equations reduce to
\begin{equation}\label{t5}
 G_i\equiv \frac{T}{2} \ln\frac{1+m_i}{1-m_i}-\sum_j J_{ij}m_j+\frac{1}{T}(1-q_2) m_i =0\quad,
\end{equation}
 where the definition $ G_i$ is introduced for later use.
As shown by the present  author \cite{tp} the $m_i$ have to satisfy the two convergence criteria
\begin{equation}\label{5}
c_1\equiv T^2-1+2q_2-q_4 >0 \,\quad,\quad c_2\equiv T^2-2q_2+2q_4 >0 \quad.
\end{equation}
Criterion    $ c_1 $ is generally accepted  and is related to the de Almeida Thouless condition \cite{at} for the SK solution. Criterion $ c_2 $ is controversial \cite{owen,abm}.

These criteria are of some importance  for the present work. They result from  an application of a theorem of Pastur \cite{pastur}, which requires the in-dependency of the variables  $ m_i $  from the  bonds $ J_{ij}$. For a Gibbs potential, indeed  these magnetizations  $ m_i $  are the free and  independent variables.  Note, that for every    thermodynamic stability analysis   one has to study  the influence  of {\it all possible  $ m_i $ values including the  $ J_{ij}$ independent  values}. This  requirement is  also essential for the integration procedure  used in this work. Thus the application of the  theorem of Pastur is justified (Compare\cite{note1} ) and $ c_2> 0 $ is a necessary (but probably not sufficient \cite{ni}) convergence condition for the expansion \cite{tp}.

Further support for validity of both criteria result from the fact, that they are necessary to prove    the  positivity of the entropy   $ S(T,{m_i})$ \cite{t00} . Simple  examples leading
to a negative entropy, if $ c_2 <0 $ , can easily be constructed (see \cite{beispiel}.) Consequences of the criteria (\ref{5})  to  the $T $-dependence of $q_2$ and  $q_4$   has already been  presented in \cite{t0}.

The present approach is related  to the studies \cite{bm,CGPM,abm} of the complexity
\beq
\Sigma(T,\Omega)=N^{-1}  \log{\cal N}(T,\Omega) \label{n}
\eeq
which describes the extensive number ${\cal N}$  of solutions of TAP equations (\ref{t5})
\beq
{\cal N}(T,\Omega)=\int_{-1}^1 \prod_i dm_i \;\delta (G_i)\; { \cal C}(\Omega)\;
  |\det \partial G_l / \partial  m_k  |\,\label{det},
\eeq
where $ G_i$ is defined in Eq.(\ref{t5}).
Constraints are  considered in the term ${ \cal C}(\Omega)$, which  is chosen in this work as
\beq
  {\cal C} =\delta( q_2-\frac{1}{N} \sum_i  m^2_i)\;\delta( q_4- \frac{1}{N}  \sum_i m^4_i)\,\delta( w- \frac{1}{2 N}  \sum_{ij} m_i J_{ij} m_j)
\eeq
and the set of constrained variables is $\Omega=\{q_2,q_4 ,w\}$. The inclusion of $ q_4 $ is new, but is essential to take into account the
restrictions due to criteria (\ref{5}).  Note that $ q_4 $ is a sum of single particle terms and the modifications due to such terms are  simple. The use of $w$ instead of the Gibbs potential is technical and simplifies the calculation.

Following the  previous work \cite{bm,CGPM,abm} and
adopting  the notation of \cite{CGPM} the calculation of the complexity leads to
\beq
\Sigma(T,\Omega)= \Sigma_0 + \log\int dm \ e^{{\cal L}(\Omega,m)} \,
\label{h}
\eeq
where 
\beq
  \Sigma_0 = -
\lambda q_2 - \mu q_4   -\Delta(1-q_2)-\frac{\Delta^2}{2\beta^2}\; 
- \frac{1}{2}\log(2\pi\beta^2q_2)+\frac{\beta^2 v^2 q_2^2} {4} \label{sig}
\eeq
and
\beq
{\cal L}(\Omega, m) =  \lambda m^2 +  \mu m^4  - \frac{[\tanh^{-1}(m)-\Delta m]^2}{2\beta^2q_2} 
- \log (1-m^2)
\label{h0}
\eeq
with $\beta=1/T $.
The variables $\lambda,\Delta,\mu ,v $ enter via the Fourier representations
of the $ \delta$-functions in the calculation. In the limit $N->\infty $  steepest decent methods are applied.   The stationary of $\Sigma(T,\Omega)$ with respect to $\lambda,\Delta,\mu ,v $ leads to
\beqa
q_2 &=&\langle\langle m^2 \rangle\rangle 
\label{h1}
\\
\beta w& = &-\frac{\beta^2}{2} v q_2^2 -q_2\Delta -\beta^2q_2(1- q_2)
\label{h2}
\\
\Delta&=&-\frac{\beta^2}{2}(1-q_2) 
+ \frac{1}{2q_2}\langle\langle m\tanh^{-1}\, (m)\rangle\rangle -\frac{\beta^2}{2}v q_2\label{h3}
\\
 q_4&=&\langle\langle m^4\rangle\rangle \label{h4}
\eeqa
 with
\beq
\langle\langle F(m) \rangle\rangle = \
\frac{1}{\int dm \ e^{{\cal L}(\Omega,m)}}\ 
\int dm \  F(m) \ e^{{\cal L}(\Omega,m)}  \label{hxx} . 
\eeq

The set  of Eqs.(\ref{h}-\ref{h3}) correspond to  the Eqs.(56-61) of  \cite{CGPM} with the
replacements $ f\rightarrow w,\phi_0 \rightarrow 0,\,
u\rightarrow v, q\rightarrow  q_2,B\rightarrow  0 $ and the additional
terms proportional to $ \mu $ resulting from the inclusion of $ q_4 $.  The apparent  differences of Eq.(\ref{sig}) and  Eq.(56) of \cite{CGPM} result from a simplification  using  Eq.(\ref{h2}). Setting  $ B=0$ corresponds to an exclusion of a non-physical solution. Eq.(\ref{h4}) is obvious and results from the stationary with
respect to $ \mu $.
For more details of the calculation and for the performed approximations  it is referred to the previous work  \cite{bm,CGPM,abm}.

 Note, however, that as long as $ q_2 $ and $ q_4 $ satisfy the criteria (\ref{5}) the value of the   determinant $ \det \partial G_k / \partial  m_l  $ in Eq.(\ref{det}) is always  positive. All previous work disregards  the modulus with not completely satisfying arguments.
 \section{applications}
 \subsection*{A. Total complexity}
\begin{figure}
\includegraphics[height=9.0cm]{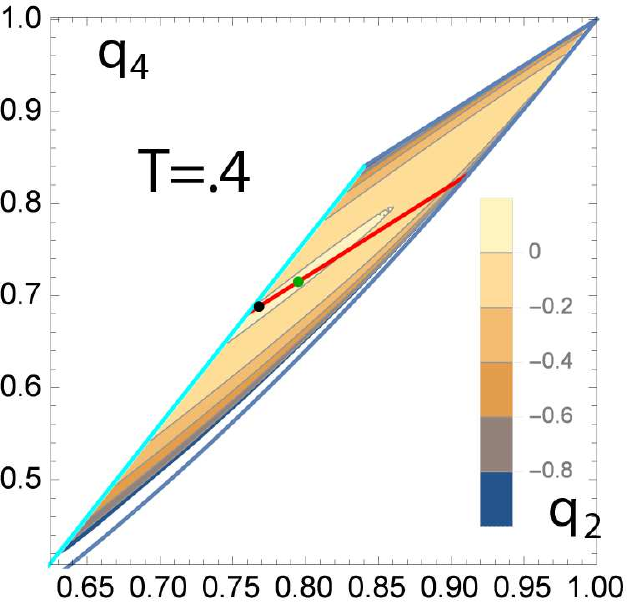}
\caption{\label{fig:1} \textbf{Contour-plot of the complexity $\Sigma(T;q_2,q_4,v=0)$ at $ T=.4\; $:} 
The cyan and the red  boundaries represent the lines  $ c_1=0$ and $ c_2=0 $. The region above the red line  is the  relevant area with $ c_2>0 $. The maximum value represents  the total complexity $\Sigma_{tot}(T=.4) $  and its position  is indicated by the green dot. The black dot indicates the position of $ g_0 $.}
\end{figure}
\begin{figure}
\includegraphics[height=12cm]{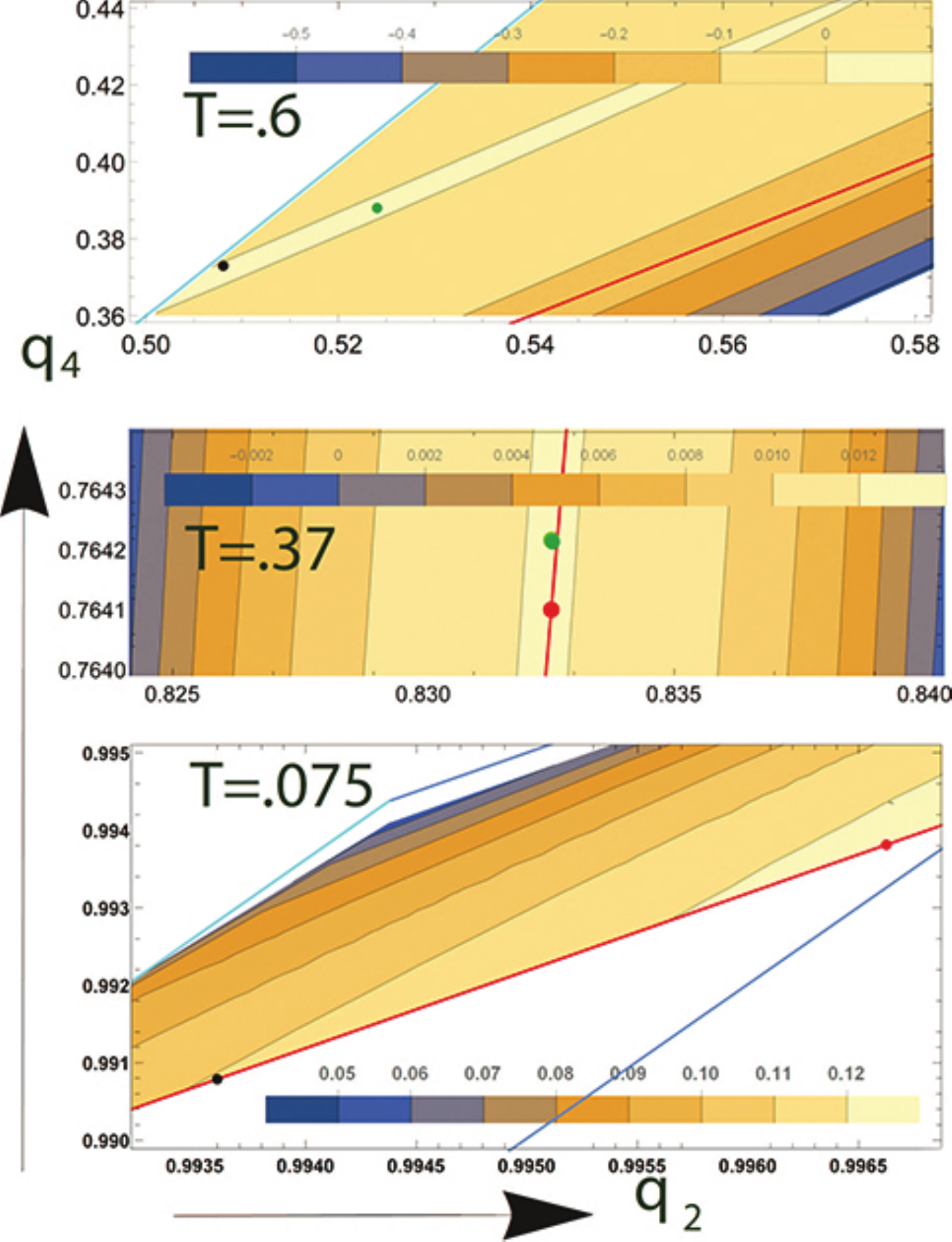}
\caption{\label{fig:2} \textbf{Contour-plot of the complexity $\Sigma(T;q_2,q_4,v=0)$ far above, near and far below the critical temperature $ T_1=.367 $:} Red  Dots mark boundary maxima and green
dots denote maxima in the interior compatible with both criteria (\ref{5}). Black dots indicate the position of $ g_0 $}
\end{figure}
  As first application  the BM work \cite{bm} for the total complexity $\Sigma_{tot} (T)$ is  reanalysed, which describes the total number of   TAP states.
  
Setting $v=0$ in Eqs.(\ref{h}-\ref{h4}) the  resulting  equations determine  the  complexity $\Sigma(T;q_2,q_4,v=0)$ for fixed values of $q_2$ and $q_4 $ . These  equations are numerical investigated for all  possible values of $ q_2,q_4 $ and for all temperatures $ T<1$.

As example  $\Sigma(T=.4;q_2,q_4,v=0)$  is plotted in the $ q_2-q_4 $ plane in FIG.\ref{fig:1} . The region of allowed  $q_2-q_4$ values is restricted by $q_2^2\leq q_4\leq q_2 $ and by the  criteria (\ref{5}). The cyan and the red  boundaries represent the lines  $ c_1=0$ and $ c_2=0 $, respectively. The physical relevant region $ c_2>0$ is above the red borderline. The region below the red line with  $ c_2<0 $ has  no physical significance.

The absolute maximum   of $\Sigma(T;q_2,q_4,v=0)$ in the  $ q_2-q_4 $ plane represents  $\Sigma_{tot} (T)$ and  can generally  be located in the interior   or on  the boundary  of the relevant region.  For temperatures  $ T\geq T_1 $ the maximum is within the region and for  $ T\leq T_1 $  the maximum is located on the boundary $ c_2=0$. The numerical value of  the critical temperature $ T_1 $ is given by 
\beq
T_1= .367  \quad .
\eeq
The coordinates of the maxima  are determined by $ \mu =0 $  and by
$  \partial \Sigma/ \partial  q_2=0 $ or by $ c_2=0 $. At $ T_1 $ the internal maximum coincides  with the boundary maximum. 

In addition to FIG.\ref{fig:1}, which gives  an overview, some details  are presented in FIG.\ref{fig:2} for $ T=.6 $ , for $ T=.37$ and for $ T=.075 $. The  internal maxima  and the boundary maxima are marked by a green and red  dots, respectively. (On the scale of
FIG. \ref{fig:1} these two minima are not separated.)
\begin{figure}
\includegraphics[height=6cm]{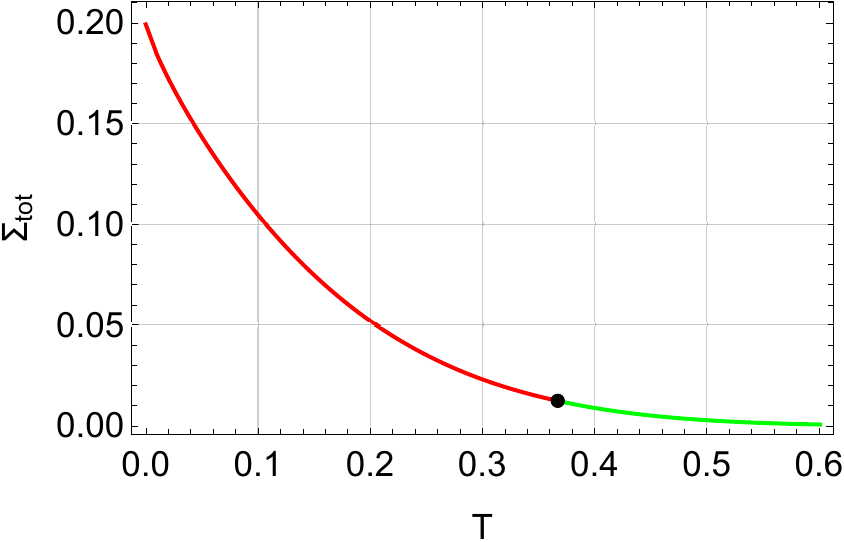}
\caption{\label{fig:3} \textbf{Total complexity  $\Sigma_{tot} (T)$ versus temperature $ T $ :} 
The red  branch result from maxima on the boundary ($T< T_1$) and the green branch result  from internal maxima ($T> T_1$).  The black dot marks the critical point $ T_ 1$. (For $ T>.6 $ the approximative expansion of  \cite{bm}  is very accurate.)}
\end{figure}
\begin{figure}
\includegraphics[height=6cm]{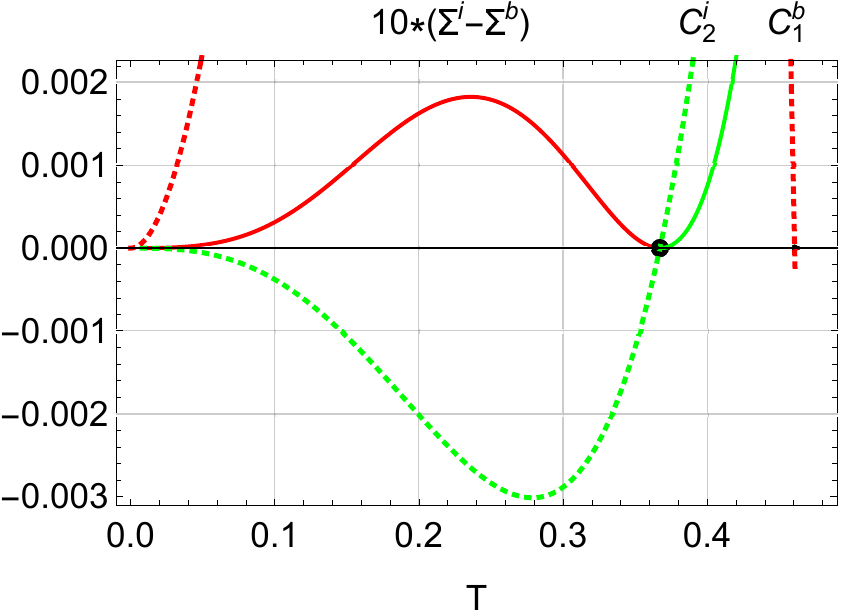}
\caption{\label{fig:4} \textbf{Separation of the extrema :} Plotted is the  difference of the internal maximum $ \Sigma^i$ to the boundary maximum 
$ \Sigma^b $  (red and green full lines). The corresponding criteria 
$ c_2^i$ and $ c_1^b $  are represented by dotted red and green  lines. The black dot marks the critical temperature $ T_1$.}
 \end{figure}

 $\Sigma_{tot} (T) $    has two branches  $\Sigma^i $ and  $\Sigma^b $ resulting from the two different maxima.  $\Sigma^i $ is identical to BM and represents the stable branch for $T>T_1$. The quantity characteristic for the transition is $ c^i_2  $, the $ c_2 $ for the internal maximum, which tends to zero for $ T \rightarrow T_1$ from above. Below $ T_1 $ criterion $ c^i_2$  is negative, the border 
maximum  is the physical one and the the branch  $\Sigma^b $ with 
 $ c^b_2=0$ is relevant. FIG.\ref{fig:3} shows the $ T $ dependence of these two branches. Both  have continuations from $ T_1 $ to the irrelevant temperatures regions. The difference of their extremal values is small (in the order of $ 10^{-4}$ ) and are plotted in  FIG.\ref{fig:4}.
 
The presented  results are new and have the important consequence  that {\it  nearly all TAP
solutions  are marginal stable for temperatures at and below $ T_1$}.
Previous claims of  marginality  \cite{muller,BM79,mod2,p03,abm,am19} are exclusively  based on $ c_1=0 $ and  differ therefore   from  the present findings, which are based on $c_2=0 $. Note in addition, that the classical BM work on the total complexity does not include any  marginality  for any temperature below T=1.

\subsection*{B. Averages}
Next some  physical interesting averages  for  the  Gibbs potential $ g$ and the energy  $ u $ per spin are calculated. According to Eqs.(\ref{t3}) and (\ref{t4})  these  averages
are given  by 
 \beq g= w -\frac{\beta}{4}(1- q_2)^2 -T\langle\langle s_0(m) \rangle\rangle \label{ggg}
\eeq
and by 
 \beq u= w -\frac{\beta}{2}(1- q_2)^2\quad.\label{uuu}
\eeq
\begin{figure}
\includegraphics[height=6.cm]{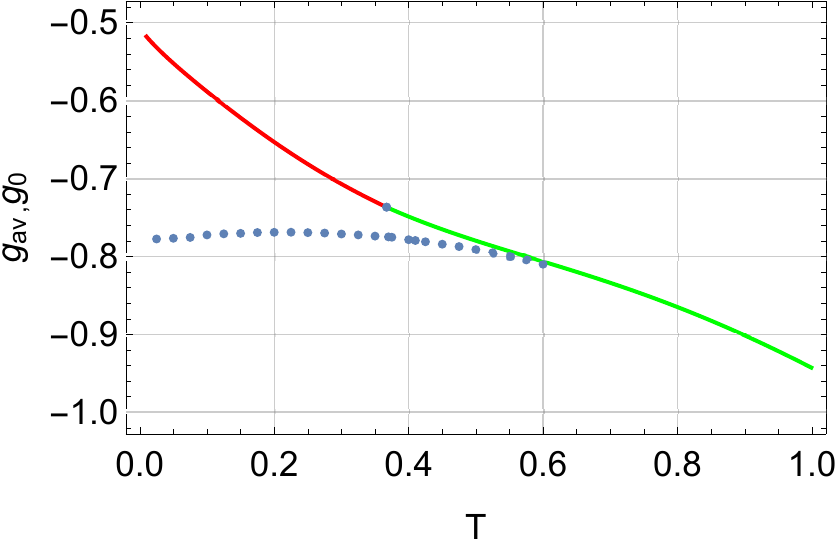}
\caption{\label{fig:5} \textbf{Gibbs potential:}  $ T$-dependence of $g_{av}$  (full lines) and $g_0$ (dots).}
\end{figure} 
\begin{figure}
\includegraphics[height=6.cm]{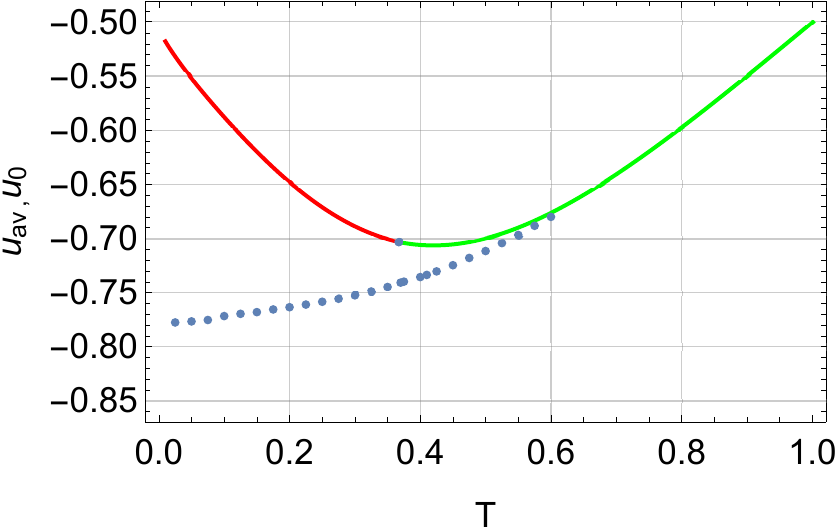}
\caption{\label{fig:6} \textbf{Energy:} $ T$-dependence of $u_{av}$  (full lines) and $u_0$ (dots).}
\end{figure}
Let us first consider the averages over all TAP states with equal weights. These 'white' averages  $ g_{av}$  and of $u_{av} $  are determined by the extremal values  of the parameters of the total complexity $\Sigma_{tot} (T)$. The temperature dependence for $ T\leq 1 $  of $ g_{av}$  and of $u_{av} $ is plotted in FIG.\ref{fig:5} and in  FIG.\ref{fig:6}, respectively.  The strange increase of $u_{av} $ with decreasing $ T $ results
from the fast increasing number  of TAP solutions with high  energies. These white averages have therefore no physical significance or any relevance for low temperatures.

There is an interesting, alternative averaging, that leads to the lowest 
value $ g_0 $ of the Gibbs  potential consistent with both, the existence condition of TAP solutions $ \Sigma \geq0  $ and  the validity of the criteria(\ref{5}).
To attack this problem the complete set of Eqs.(\ref{h}-\ref{h4},\ref{ggg},\ref{uuu})  is needed.
Keeping $  q_2, q_4 $ and $ v $   constant  the  parameters $\lambda,\Delta,$ and $\mu$ are determined numerically 
with Eqs.(\ref{h1},\ref{h3},\ref{h4}). Repeating this procedure for all possible values of  $  q_2, q_4 $ and  $ v $  
the dependence of the  complexity $ \Sigma $ , of $w $ and of the Gibbs potential $g$   on these quantities is obtained according  to 
 Eqs.(\ref{sig},\ref{h2},\ref{ggg}). Finally the  minimum $ g_0 $ of the Gibbs potential  is determined in the region of the allowed values  of $ q_2$ ,  $ q_4$ and $ \Sigma $. The findings are a vanishing complexity $ \Sigma= 0 $ for all temperatures, which ensures at least the presence of one TAP state in the thermodynamic limit $ N \rightarrow\infty $.

The resulting $ g_0 $ and the corresponding  energy $ u_0 $ are plotted  in FIG.\ref{fig:5} and in  FIG.\ref{fig:6},
respectively.  Both quantities  exhibits  the expected temperature dependence and are similar to the  results of the
replica approach  \cite{andrea}.  This is remarkable as both approaches use different averages. A second interesting
point is the fact that  the location of $ g_0$ is again on the boundary $ c_2=0 $ for $ T<T_1$ . (Compare FIG.\ref{fig:2}.)

\subsection*{C. Postulated marginality}

 Previous numerical  investigations   \cite{BM79,mod2,p03,abm,am19} claim  marginal metastable states based on  $ c_1\rightarrow 0 $ in the thermodynamic limit. Therefore  $ c_1=0 $ is  postulated a priory in this  subsection  and the resulting consequences for the present approach    are worked out.
 
\begin{figure}
\includegraphics[height=7cm]{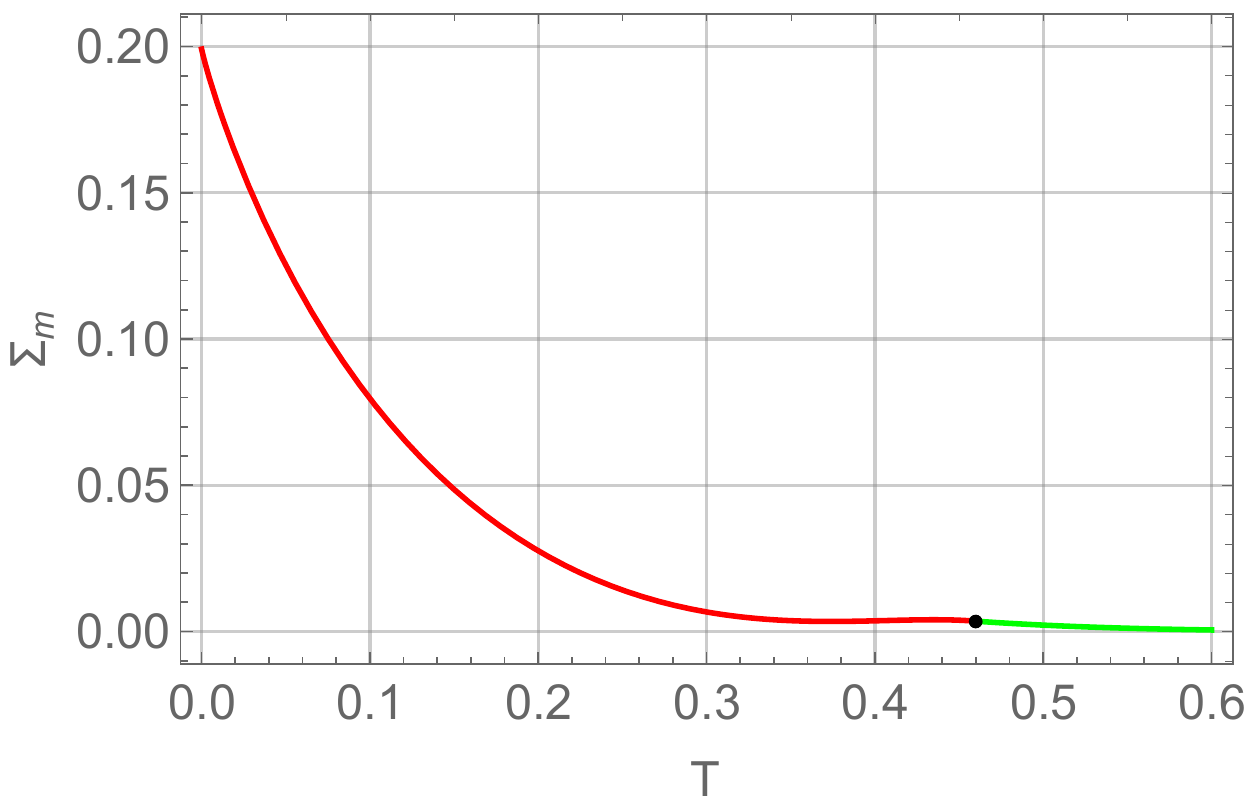}
\caption{\label{fig:77} \textbf{ Complexity  $\Sigma_{m} $ versus temperature $ T $ :} The red  branch result from maxima on the boundary ($T< T_{av}^m$) and the green branch result  from internal maxima ($T>T_{av}^m $).  The black dot marks the critical temperature $ T_{av}^m$.}
\end{figure}

The first quantity of interest is the complexity $\Sigma_{m} $,
which determines the total number  of TAP states with the   
constraint $ c_1=0 $ . This complexity  $\Sigma_{m} $ is given by the
absolute  maximum   of  $\Sigma(T;q_2,q_4=2q_2-1+ T^2,v=0)$ , as function of $q_2.$ Again  two temperature regimes exist, which are separated  by a critical temperature $ T_{av}^m =.459 $ . For $ T> T_{av}^m $  the maximum is located within the allowed $q_2$ - interval $ (1-1.5\; T^2) <q_2<1 $  and for  $ T\leq T_{av}^m$  the maximum sticks at the endpoint $ q_2=(1-1.5 \; T^2) $ . 
The numerical results for $\Sigma_{m} $ are consistent to $ 0 \leq \Sigma_{m}\leq\Sigma_{tot} $ for all temperatures, which  ensures the existence of at least one
TAP state with the postulated marginality. Note that this last conclusion is a priory not obvious. FIG.\ref{fig:77} shows the temperature dependence of $\Sigma_{m} $  for $ T<.6 $ .

\begin{figure}
\includegraphics[height=7.cm]{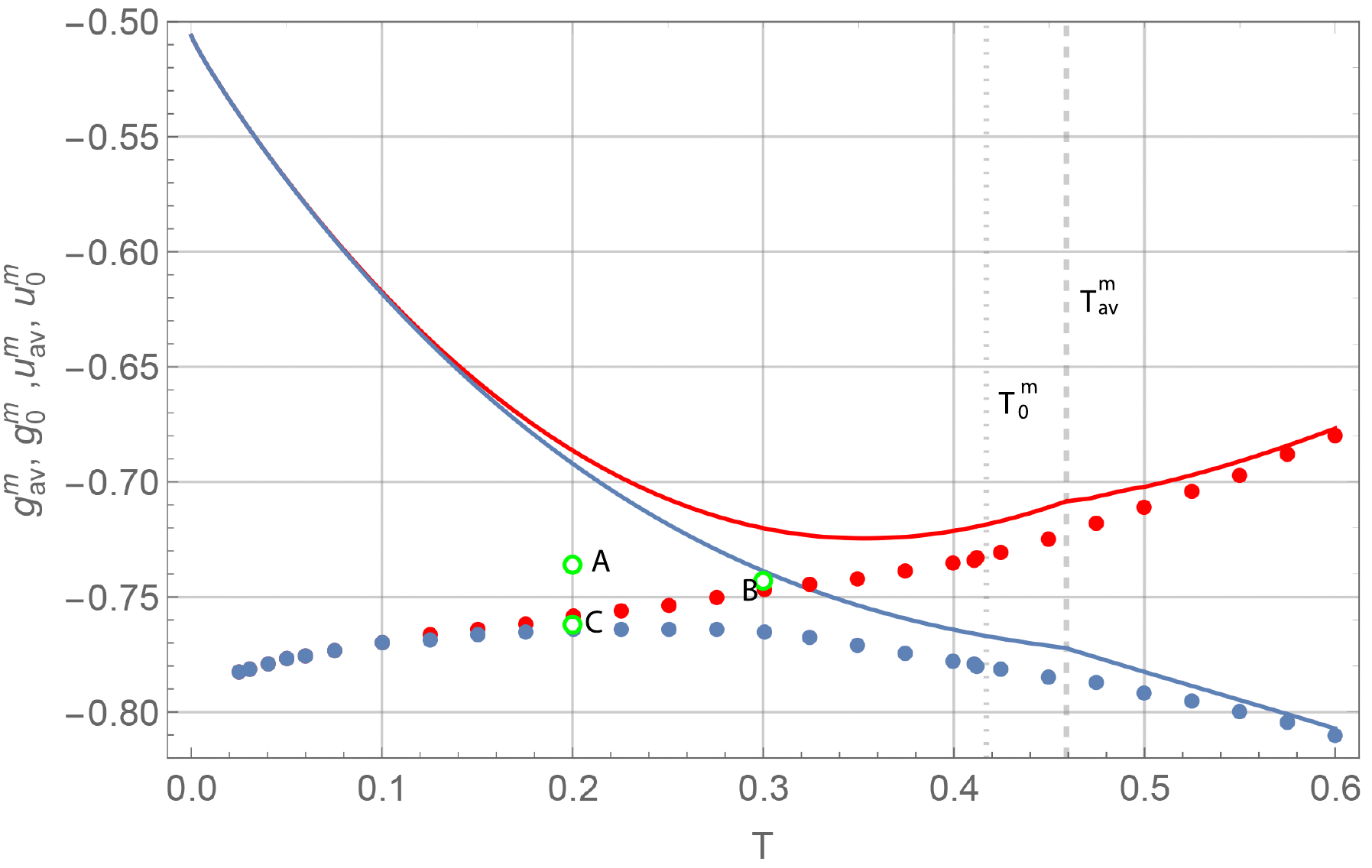}
\caption{\label{fig:88} \textbf{ Gibbs Potential and Energy:} $ T$-dependence of $g_{av}^m$  (full, blue line), $u_{av}^m$ (full, red line), $g_{0}^m$  (blue dots) and $u_{0}^m$ (red dots). The green points A, B and C represent published values of the Gibbs potential according to
\cite{abm}, to \cite{am19} and to \cite{p03}, respectively.}
\end{figure}

Analogue to subsection B. the set of extremal parameters of the maximum  determine the averages performed with all TAP states satisfying  $ c_1=0$. Together with Eqs.(\ref{ggg}) and (\ref{uuu}) this leads directly to the averages  of the Gibbs potential $ g_{av}^m $ and the energy $ u_{av}^m $ performed with these states . Both quantities $ g_{av}^m $ and  $ u_{av}^m $
are plotted in FIG.\ref{fig:88}. The results $ g_{av}^m $ and $ u_{av}^m$ are not very useful for low temperatures similar to the above findings for $ g_{av} $ and $ u_{av} $.

The lowest value of the Gibbs potential $ g_0^m $ consistent with $ c_1=0\;,\;c_2\geq 0$ and $ \Sigma\geq 0 $ is again determined analogue to procedure of subsection B. As before a vanishing complexity  and two temperature regions are found with a sticking below a different critical temperature  $ T_{0}^m =.417 $. The temperature dependence of $ g_0^m $ together with the corresponding energy $ u_0^m $ is plotted in FIG.\ref{fig:88}. 

For  $ T=.2 $  the numerical value is for the Gibbs potential is $ g_0^m =-.7644 $.
This value is in remarkable agreement  with the value of $ \hat{g}_0^m= -.07619 $  found by the dynamical investigation \cite{p03} with $ c_1->0 $ and with  a vanishing complexity. 
The  result $ \hat{g}_0^m $ is added to  FIG.\ref{fig:88} as Point $C$  together with the Points $A$ and $B$  found by  further investigations \cite{abm,am19}. Even  
though  $ c_1->0 $ is satisfied no results for the complexity are available for the points $A$ and $B$. Nevertheless these
results are compatible with the present work.
 \section{Conclusion}
 The presented  investigation  of the  complexity  is based  the  inclusion of $ q_4 $ to the set of  constrained variables. This extension together with a strict regard  of the validity criteria for the TAP equations leads to new results in 
low  temperatures regime.  Numerically the differences to the existing approaches are
rather small, the interpretations  and conclusions, however, differ considerably.
At and below a critical temperature $ T_1 $ {\it  nearly all TAP states  are marginal stable with $ c_2=0$}, a property  not found in previous theories on the total complexity.
Marginal stability implies  a vanishing eigenvalue of  the Hessian and  the divergence of a mode of the susceptibility  matrix \cite{tp}. The system is critical for all  temperatures below $T_1 $ and shows critical slowing down effects.
 
In addition to this findings consequences for averages over all metastable  TAP states and averages over states with the lowest value of the Gibbs potential have
been worked out for all temperatures. Moreover the influence  of a different kind
of marginality  $ c_1=0$ , as found by supplementary numerical investigations \cite{BM79,mod2,p03,abm,am19}, has been worked out.
\begin{acknowledgments}
I acknowledge helpful discussions with Nicola Kistler and Jan Plefka
\end{acknowledgments}



\begin{references}
\bibitem {note1}
  Owen \cite{owen} investigates a different question, as he conciders   $ m_i$ , which depend on the $ J_{ij}$ via the TAP equations and the Pastur Theorem can not be applied. There is  no  disagreement to  \cite{tp}, which investigates  the Gibbs potential with free and independent  $ m_i $ .
Moreover, it is generally  impossible to conclude 
anything on the convergence of a series from a partial re-summation, as done in \cite{owen}.
\bibitem{beispiel}
Consider  a special set with $m_i^2=1-T $, which implies $ c_1=0 $ and $c_2= T(3T-2) $. For $ T<2/3 $ condition $ c_2$ is violated.
The  entropy  is given by $ S/N = 
 s_0(\sqrt{1-T})-1/4 $, which is negative for all temperatures smaller than $ 0.256 $.
\end{references}
\end{document}